\documentclass[aps,pre,reprint,groupedaddress]{revtex4-1}
\usepackage{amssymb}
\usepackage{amsmath}
\usepackage{amsfonts}
\usepackage{graphicx}

\begin{document}

\title{Seismic hemispheric asymmetry induced by Earth's inner core decentering}
\author{C\u{a}lin Vamo\c{s}}
\email{cvamos@ictp.acad.ro}
\affiliation{ ``T. Popoviciu'' Institute
of Numerical Analysis, Romanian Academy, P.O. Box 68, 400110
Cluj-Napoca, Romania}
\author{Nicolae Suciu}
\email{suciu@am.uni-erlangen.de}
\affiliation{Department of
Mathematics, Friedrich-Alexander University of Erlangen-Nuremberg,
Martensstr. 3, 91058 Erlangen, Germany}

\begin{abstract}
In a first approximation the Earth's interior has an isotropic
structure with a spherical symmetry. Over the last decades the
geophysical observations have revealed, at different spatial scales,
the existence of several perturbations from this basic structure.
Some of them are situated in the neighborhood of the inner core
boundary (ICB). One of the best documented perturbations is the
asymmetry at the top of the inner core (ATIC) characterized by
faster seismic wave velocity in the eastern hemisphere than in the
western hemisphere. All existing explanations are based on a
hemispheric variation of the material properties near ICB inside the
inner core. Using numerical simulations of the seismic ray
propagation, we show that the ATIC can be explained as well by the
displacement of the inner core towards east in the equatorial plane
tens of kilometers from the Earth's center, without modifying the
spherical symmetry in the upper inner core. The hypothesis of a
displaced inner core is also sustained by other observed hemispheric
asymmetries at the top of the inner core and at the bottom of the
outer core. A displaced inner core would have major implications for
many mechanical, thermal, and magnetic phenomena in the Earth's
interior.
\end{abstract}

\maketitle


\section{Introduction}
Over the last decades the geophysical observations have revealed the
existence of several perturbations of the Earth's interior from an
isotropic structure with a spherical symmetry
\cite{Poupinetetal1983,Morellietal1986,GarciaandSouriau2000,
OuzounisandCreager2001,NiuandWen2001,WenandNiu2002,SumitaandOlson2002,
Yuetal2005,YuandWen2006,Souriau2007,Aubertetal2008,SunandSong2008,
Monnereauetal2010,Deussetal2010,RohrbachandSchmidt2011,Waszeketal2011}.
One of the best documented perturbations is the asymmetry at the top
of the inner core (ATIC) characterized by faster seismic wave
velocity in the eastern hemisphere than in the western hemisphere
\cite{GarciaandSouriau2000,OuzounisandCreager2001,
NiuandWen2001,WenandNiu2002,Yuetal2005,YuandWen2006,Waszeketal2011}.
All existing explanations assume a hemispheric variation of the
material properties near the inner core boundary (ICB) inside the
inner core
\cite{NiuandWen2001,WenandNiu2002,Yuetal2005,YuandWen2006,
Alboussiereetal2010,Waszeketal2011}. We show that the ATIC can be
explained as well by the displacement of the inner core towards east
in the equatorial plane tens of kilometers from the Earth's center,
without modifying the spherical symmetry in the upper inner core.

The Earth's inner core is a rigid sphere surrounded by the fluid
with smaller density of the outer core. Therefore, in the mechanical
equilibrium with respect to the gravitational and hydrostatic
forces, its center of mass coincides with the Earth's center. The
displacements from the equilibrium position have been attributed to
harmonic oscillations with amplitudes of at most 0.5 m and periods
of 4-8 hr \cite{WonandKuo1973,BuffettandGoertz1995}. Movements over
tens of kilometers would imply the presence of some forces balancing
the gravitational one. They could originate from the interaction of
the inner core with the flow in the outer core and with the
terrestrial magnetic field
\cite{Aubertetal2008,BuffettandBloxham2000,BuffettandGlatzmaier2000}.
The angular momentum conservation would cause a global scale flow in
the outer core, influencing the generation of the geomagnetic field.
Therefore, the time scale of the large amplitude movements of the
inner core may coincide with that of the variations of the magnetic
field, being larger than thousands of years
\cite{Aubertetal2008,BuffettandGlatzmaier2000,Buffett2010}. The
angular momentum conservation would also imply changes in angular
velocity and rotation axis of the inner core, as several seismic
studies suggest
\cite{Aubertetal2008,Suetal1996,DumberryandMound2010}. In this
letter we leave aside the dynamical aspects of the inner core
movement and focus on an imaging interpretation of the seismic data
in order to obtain information on the actual position of the inner
core.

\section{Seismic rays for decentered inner core}
\begin{figure*}
\includegraphics{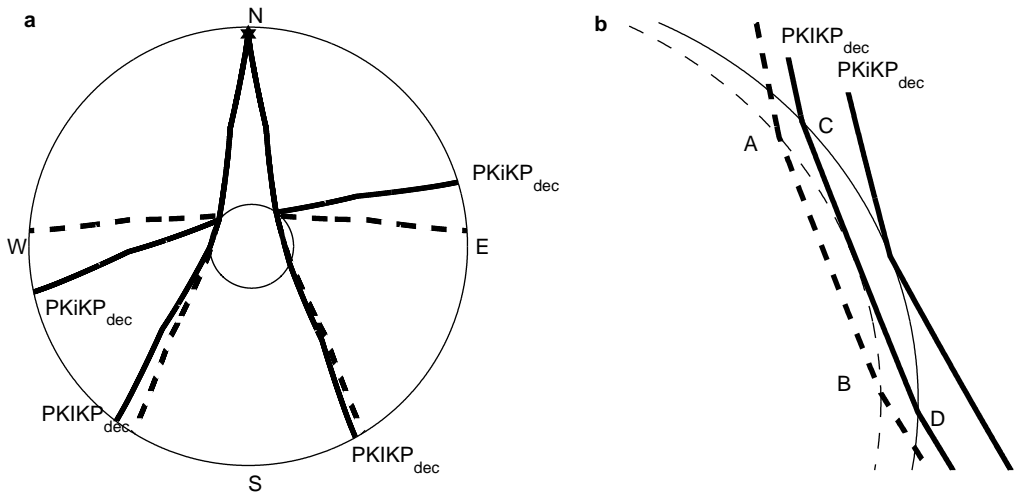}
\caption{\label{fig1} Propagation of the PKIKP seismic phase for
decentered inner core. (a) The paths of the PKIKP$_{\mathrm{dec}}$
and PKiKP$_{\mathrm{dec}}$ rays (thick continuous lines) when the
inner core (the small circle) is displaced from the Earth's center
with 100 km towards east. The propagation plane of the seismic ray
contains the centers of both Earth and decentered inner core and in
this case it is not modified by refraction on ICB. The dashed lines
represent the paths of the same phases for the centered inner core.
All the represented seismic rays have the same initial incidence
angle; hence they are identical until the incidence with the ICB.
After that, the paths for the centered inner core are symmetric,
while PKIKP$_{\mathrm{dec}}$ and PKiKP$_{\mathrm{dec}}$ rays clearly
exhibit an east-west asymmetry. (b) The path inside the eastern
hemisphere of the decentered inner core of the
PKIKP$_{\mathrm{dec}}$ ray presented in panel a. The PKIKP (thick
dashed line) and PKiKP$_{\mathrm{dec}}$ rays have not the same
initial incidence angle as in panel a, but they emerge at the same
point on the Earth's surface as the PKIKP$_{\mathrm{dec}}$ ray. The
circular arcs are the ICB when the inner core is decentered
(continuous line) and when it is centered (dashed line).}
\end{figure*}

ATIC manifests itself predominantly in the residuals of the
differential travel time of the PKIKP and PKiKP seismic phases
\cite{NiuandWen2001,WenandNiu2002,Yuetal2005,YuandWen2006,Waszeketal2011}.
They both travel through almost the same regions of the crust,
mantle, and outer core. After that, the PKiKP phase reflects off the
ICB, while the PKIKP phase refracts twice on ICB propagating inside
the inner core. If the inner core is displaced from the Earth's
center, then the paths of the two seismic phases change after
reflection and refraction on its boundary (Fig.~\ref{fig1}a). We
denote by PKIKP$_{\mathrm{dec}}$ and PKiKP$_{\mathrm{dec}}$ the
paths modified by the decentered inner core. Unlike the paths for
the centered inner core, their propagation plane changes at
reflection or refraction on ICB.

We compute the differential travel time $\Delta t$ by subtracting
the travel time of the PKIKP phase from the travel time of the PKiKP
phase with the same epicentral distance. We denote by $\Delta t_0$
the differential travel time computed from the velocity profile of a
1D reference seismic model. The observational data show that the
residuals $\Delta t-\Delta t_0$ are positive in the eastern
hemisphere and negative in the western hemisphere
\cite{NiuandWen2001,WenandNiu2002,Yuetal2005,YuandWen2006,Waszeketal2011}.
Under the hypothesis that the inner core is concentric with the
Earth, this asymmetry is explained by the greater (smaller) seismic
wave velocity at the top of the inner core in the eastern (western)
hemisphere than the velocity of the 1D reference models.

When the inner core is decentered, the epicentral distance and the
travel time of the PKIKP$_{\mathrm{dec}}$ and PKiKP$_{\mathrm{dec}}$
phases depend on the initial propagation plane of the seismic ray
and on the earthquake focus location. Because of the shifted
position of the inner core, the total length, from focus to exit
point, of the PKIKP$_{\mathrm{dec}}$ ray is smaller than that of
PKIKP corresponding to the centered inner core (Fig.~\ref{fig1}b).
The length of the reflected ray PKiKP$_{\mathrm{dec}}$ also
decreases by approximately the same amount. In the diametrically
opposite region of the inner core both lengths have approximately
the same increase. These changes in path lengths for the pairs of
reflected and refracted phases do not change the differential travel
time $\Delta t_{\mathrm{dec}}$.

There is another geometric effect which modifies the differential
travel time $\Delta t_{\mathrm{dec}}$. The segment CD of the seismic
ray within the decentered inner core is longer than the segment AB
for the centered inner core (Fig.~\ref{fig1}b). Because the velocity
in the inner core is larger than in the outer core, the travel time
of the PKIKP$_{\mathrm{dec}}$ phase has an additional decrease, the
differential travel time $\Delta t_{\mathrm{dec}}$ increases, and
the residual $\Delta t_{\mathrm{dec}}-\Delta t_0$ is positive. In
the diametrically opposite region of the inner core the distance CD
is smaller than AB and the residual is negative, resulting in a
hemispheric asymmetry. Hence, the asymmetry of the residuals of the
differential travel time can be explained by the variation of the
PKIKP$_{\mathrm{dec}}$ ray paths in the decentered inner core
without modifying the seismic velocities.

\section{Numerical results}
\begin{figure}
\includegraphics{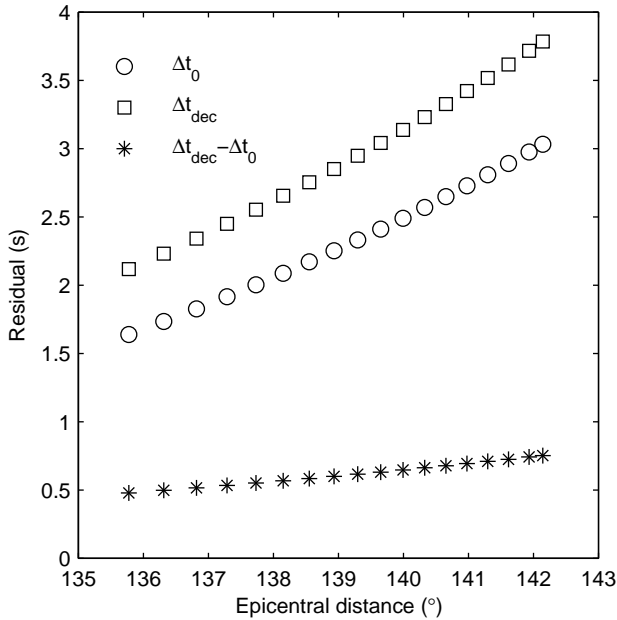}
\caption{\label{fig2} Numerically computed residuals for a
decentered inner core. The differential travel times $\Delta
t_{\mathrm{dec}}$ are computed for the PKIKP$_{\mathrm{dec}}$ ray in
the eastern hemisphere presented in Fig.~\ref{fig1}b. We compute the
reference differential travel time $\Delta t_{0}$ for the seismic
model ak135 \cite{Kennettetal1995}. The variation range of the
epicentral distance corresponds to the range 39-90 km of the turning
point depth of the PKIKP ray in the centered inner core analyzed in
\cite{Waszeketal2011}.}
\end{figure}

The numerically computed residuals (Fig.~\ref{fig2}) are
quantitatively comparable with those observed
\cite{NiuandWen2001,WenandNiu2002,Yuetal2005,YuandWen2006,Waszeketal2011}
showing that displacements of the inner core over distances up to
100 km could explain ATIC. In the computations we assume that the
velocity profile in the decentered inner core is that of the model
ak135 \cite{Kennettetal1995}. Outside the inner core we also use the
model ak135. It is linearly extrapolated to the points of the outer
core situated at a smaller distance from the Earth's center than the
inner core radius. Each of the two regions is divided into spherical
layers with constant velocity of 1 km maximum thickness. The
boundaries of the spherical layers also contain all the reference
levels of the ak135 model.  Therefore the numerical seismic rays
consist of straight segments satisfying the refraction and
reflection laws at the boundaries of the spherical layers. With the
increase of the turning point depth (epicentral distance) the
positive residuals in the eastern hemisphere increase because the
segment CD increases (see Fig.~\ref{fig1}b).

In order to ascertain if a decentered inner core could explain ATIC,
we should compare the longitudinal repartition of the residuals
obtained by numerical simulations with those reported in
\cite{Waszeketal2011}, the most extensive and accurate PKiKP -PKIKP
study to date. The positions of the simulated earthquake sources are
given by the spherical coordinate angles with respect to the axis
defined by the centers of the Earth and the decentered inner core.
These angles vary by steps of 10$^{\circ}$. The seismic rays are
emitted from each focus in planes making with each other angles of
10$^{\circ}$. We compute the residuals for the minimum depth below
ICB (39 km) of the turning point of the PKIKP$_{\mathrm{dec}}$ ray
for which observational data are available \cite{Waszeketal2011}.

\begin{figure*}
\includegraphics{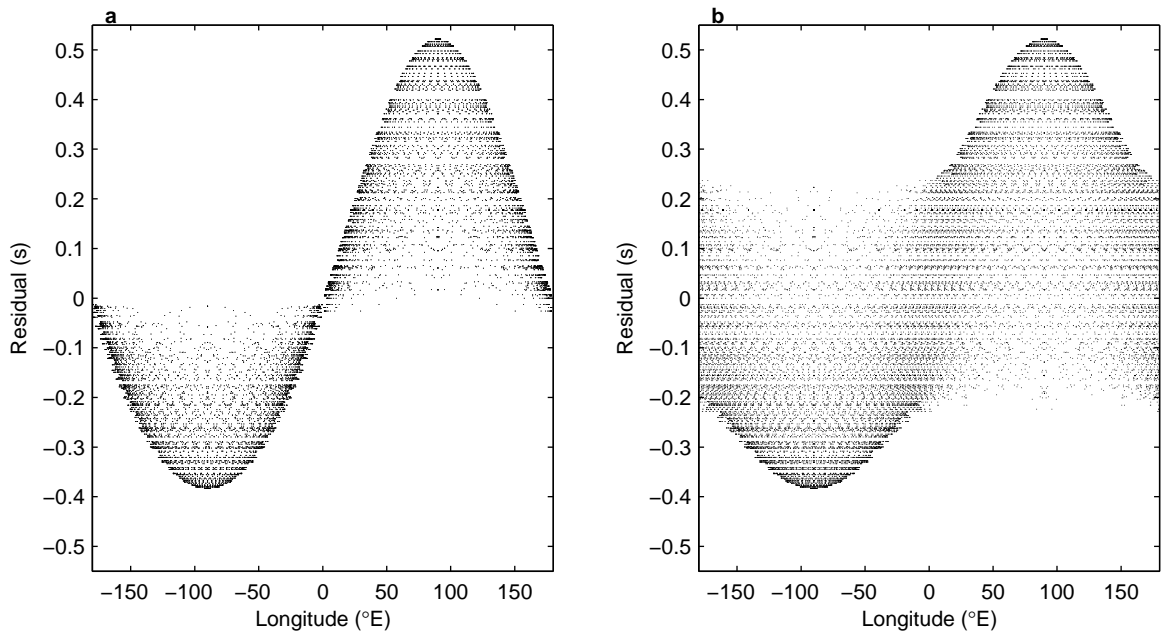}
\caption{\label{fig3} Longitudinal distribution of the residuals
obtained by numerical simulation. The inner core is displaced with
100 km towards 90$^{\circ}$ east longitude in the equatorial plane
(a) and along a direction making an angle of 30$^{\circ}$ with the
equatorial plane (b). On the abscissa we have the longitude of the
turning point of the PKIKP ray refracted by a centered inner core
which emerges at the same point on the Earth's surface as the
PKIKP$_{\mathrm{dec}}$ ray refracted by a decentered inner core.}
 \end{figure*}

We computed the residuals for a displacement of the inner core of
100 km towards 90$^{\circ}$ east longitude (Fig.~\ref{fig3}). If the
displacement is in the equatorial plane, then the positive residuals
are confined within the eastern hemisphere and the negative ones
within the western hemisphere (Fig.~\ref{fig3}a). The position of
the boundary between the hemispheres with positive and negative
residuals rotates with the angle between the plane of the
90$^{\circ}$  east meridian and the direction defined by the centers
of the Earth and the decentered inner core. If the inner core is
displaced outside the equatorial plane, the separation of the
positive and negative residuals is not so definite
(Fig.~\ref{fig3}b).

In the observational data, the positive and negative residuals are
sharply separated \cite{Waszeketal2011} corresponding to a
displacement of the inner core in equatorial plane as in
Fig.~\ref{fig3}a. The boundary between them does not coincide with
that between the eastern and western hemispheres, being shifted
towards east by approximately 20$^{\circ}$. All these indicate a
displacement of the inner core with tens of kilometers in equatorial
plane towards 110$^{\circ}$  east longitude.

In comparing the results of the numerical simulations with the
observational data we have to take into account the simplifying
hypotheses of the numerical simulation as well as the observational
errors. For instance, we use the velocities of the ak135 model
obtained under the hypothesis that the inner core is centered. Or
the observational differential travel times PKiKP-PKIKP have
fluctuations around the values derived from the model ak135 with an
amplitude of 0.5 s \cite{Kennettetal1995}, comparable with the
values associated with ATIC \cite{Waszeketal2011}. That is why the
exact longitude separating the positive and negative observed
residuals and its variation with the turning point depth of the
seismic rays cannot be determined precisely. The observational data
suggest an eastward shift of the hemisphere boundary with increasing
depth, while the numerical simulation shows that it does not vary
with the depth.

\section{Discussion and Conclusion}
There are other seismic phenomena with east-west asymmetry explained
by a decentered inner core, but none of them has a complete
observational description of the longitudinal variation. For
instance, ATIC is associated to a hemispheric asymmetry of the
seismic waves attenuation
\cite{WenandNiu2002,LiandCormier2002,CaoandRomanowicz2004,
OreshinandVinnik2004,YuandWen2006,Souriau2007,Iritanietal2010} which
seems to be confined to the uppermost inner core
\cite{LiandCormier2002,Souriau2007}. Existing explanations of the
attenuation asymmetry require a trade-off between attenuation and
velocity structures in the inner core and velocity structure at the
bottom of outer core \cite{WenandNiu2002,YuandWen2006}. If the inner
core is decentered, the PKIKP$_{\mathrm{dec}}$ phase propagates in
the eastern hemisphere over a longer distance inside the inner core
(segment CD in Fig.~\ref{fig1}b) than in the western hemisphere.
Since the quality factor $Q$ is two orders of magnitude larger in
the outer core than in the inner core \cite{Kennettetal1995}, the
attenuation $Q^{-1}$ in the eastern hemisphere is larger than in the
western hemisphere.

Another example of hemispheric asymmetry is the observation that
PKiKP phases sampling the eastern hemisphere arrive by about 0.9 s
earlier than those sampling the western hemisphere
\cite{Yuetal2005}. When the inner core is displaced eastwards, the
length of the PKiKP$_{\mathrm{dec}}$ ray in the eastern hemisphere
is smaller than in the western hemisphere (Fig.~\ref{fig1}b). Our
numerical simulations reproduce fairly well the observed PKiKP
travel time hemispheric asymmetry.

The displacement of 100 km of the inner core should produce
noticeable effects in the neighborhood of the ICB at the bottom of
the outer core as well. If $r_c$ is the radius of the inner core,
then the distances from the Earth's center to the ICB would vary
between $r_c-100$ and $r_c+100$. If the inner core were decentered
but the seismic model assumes that it is centered, the
interpretation of the seismic observations would have larger errors
in the spherical layer of 200 km containing the ICB than in other
regions of the Earth's interior. Indeed, the reference 1D seismic
models are different from each other over a thickness of roughly 200
km above ICB \cite{Yuetal2005,YuandWen2006,Souriau2007}.

A decentered inner core should cause hemispheric asymmetries in the
repartition of the physical quantities above ICB. Therefore our
numerical model is only a first order approximation and other
consequences of the inner core displacement on the structure of the
Earth's interior have to be analyzed.

\end{document}